\documentclass{rspublic}
\usepackage[dvips]{graphicx}
\usepackage{epsfig}

\begin{document}

\title[]{Discrete-time ratchets, the Fokker-Planck equation and Parrondo's paradox}

\author[]{ P. Amengual$^{1,2}$, A. Allison$^{2}$, R. Toral$^{1}$ and D. Abbott$^{2}$}
\affiliation{$^1$ Instituto Mediterr\'aneo de Estudios Avanzados, IMEDEA (CSIC-UIB),\\
Ed. Mateu Orfila, Campus UIB, E-07122 Palma de Mallorca, Spain\\
$^2$ Centre for Biomedical Engineering (CBME)
and \\ Department of Electrical and Electronic Engineering,\\
The University of Adelaide, SA 5005, Australia}\label{firstpage}

\maketitle

\begin{abstract}{Parrondo's paradox; Fokker-Planck equation; Brownian ratchet.}
Parrondo's games manifest the apparent paradox where losing
strategies can be combined to win and have generated significant
multidisciplinary interest in the literature. Here we review two
recent approaches, based on the Fokker-Planck equation, that
rigorously establish the connection between Parrondo's games and a
physical model known as the flashing Brownian ratchet. This gives
rise to a new set of Parrondo's games, of which the original games
are a special case. For the first time, we perform a complete
analysis of the new games via a discrete-time Markov chain (DTMC)
analysis, producing winning rate equations and an exploration of
the parameter space where the paradoxical behaviour occurs.
\end{abstract}

\section{Introduction}

In many physical and biological systems, combining processes may
lead to counter-intuitive dynamics. For example, in control
theory, the combination of two unstable systems can cause them to
become stable (Allison  \& Abbott 2001a). In the theory of
granular flow, drift can occur in a counter-intuitive direction
(Rosato \emph{et al.} 1987; Kestenbaum 1997). Also the switching
between two transient diffusion processes in random media can form
a positive recurrent process (Pinsky \& Scheutzow 1992). Other
interesting phenomena where physical processes drift in a
counter-intuitive direction can be found (see for example Adjari
\& Prost 1993; Maslov \& Zhang 1998; Westerhoff \emph{et al.}
1986; Key 1987; Abbott 2001).

The Parrondo's paradox (Harmer \& Abbott 1999a,b; Harmer \emph{et
al.} 2000) is based on the combination of two negatively biased
games -- losing games -- which when combined give rise to a
positively biased game, that is, we obtain a winning game. This
paradox is a translation of the physical model of the Brownian
ratchet into game-theoretic terms. These games were first devised
in 1996 by the Spanish physicist Juan M. R. Parrondo, who
presented them in unpublished form in Torino, Italy (Parrondo
1996). They served as a pedagogical illustration of the flashing
ratchet, where directed motion is obtained from the random or
periodic alternation of two relaxation potentials acting on a
Brownian particle, none of which individually produce any net flux
(see Reimann 2002 for a complete review on ratchets).

These games have attracted much interest in other fields, for
example quantum information theory (Abbott \emph{et al.} 2002;
Flitney \emph{et al.} 2002; Meyer \& Blumer 2002a; Lee \emph{et
al}. 2002a), control theory (Kocarev \& Tasev 2002; Dinis \&
Parrondo 2002), Ising systems (Moraal 2000), pattern formation
(Buceta \emph{et al.} 2002a,b; Buceta \& Lindenberg 2002),
stochastic resonance (Allison \& Abbott 2001b), random walks and
diffusions (Cleuren \& Van den Broeck 2002; Key \emph{et al.}
2002; Kinderlehrer \& Kowalczyk 2002; Percus \& Percus 2002; Pyke
2002), economics (Boman \emph{et al.} 2002), molecular motors in
biology (Ait-Haddou \& Herzog 2002; Heath \emph{et al.} 2002) and
biogenesis (Davies 2001). They have also been considered as
quasi-birth-death processes (Latouche \& Taylor 2003) and lattice
gas automata (Meyer \& Blumer 2002b).

Parrondo's two original games are as follows. Game A is a simple
tossing coin game, where a player increases (decreases) his
capital in one unit if heads (tails) show up. The probability of
winning is denoted by $p$ and the probability of losing is $1-p$.

Game B is a capital dependent game, where the probability of
winning depends upon the actual capital of the player, modulo a
given integer $M$. Therefore if the capital is $i$ the probability
of winning $\pi_i$ is taken from the set
$\{\pi_0,\pi_1,\dots,\pi_{M-1}\}$ as $\pi_i=\pi_{i\,{\rm mod}\,
M}$. In the original version of game B, the number $M$ is set
equal to three and the probability of winning can take only two
values, $p_1,\,p_2$, according to whether the capital of the
player is a multiple of three or not, respectively. Using the
previous notation we have $p_1\equiv\pi_0$,
$p_2\equiv\pi_1=\pi_2$. The numerical values corresponding to the
original Parrondo's games (Harmer \& Abbott 1999a) are:
\begin{equation}\label{probaborig}
   \left\{ \begin{array}{cc}
    p=&\frac{1}{2}-\epsilon,\\
    p_1=&\frac{1}{10}-\epsilon,\\
    p_2=&\frac{3}{4}-\epsilon,
    \end{array}\right.
\end{equation}
where  $\epsilon$ is a small biasing parameter introduced to
control the three probabilities.

Although the original game B was based on a modulo rule, there are
other versions of Parrondo's games where this rule has been
replaced by a history dependent rule (Parrondo \emph{et al}.
2000); also combinations of two history dependent games are
considered (Kay \& Johnson 2002). Instead of a random alternation,
also chaotic alternation between the games has been studied (Arena
\emph{et al.} 2003). Effects of cooperation between players have
also been considered in Parrondo's games, where the probabilities
of game B depend on the actual state of the neighbours of the
player (Toral 2001), also a redistribution of capital between the
players has been considered (Toral 2002). Other variations of
collective games have recently appeared (Mihailovi\'c \&
Rajkovi\'c 2003a,b). For a full review of Parrondo's paradox see
Harmer \& Abbott 2002.

Games A and B appearing in the Parrondo's paradox can be thought
of as diffusion processes under the action of a external
potential. However, they do not have the more general form of a
natural diffusion proces, because the capital will always change
with every game played, whereas in the most general diffusion
process a particle can either move up or down or remain in the
same position at a given time. In this article we present a new
version of Parrondo's games, where a new transition probability is
taken into account. We introduce a \emph{self-transition}
probability, that is, the capital of the player now can remain the
same after a game played with a probability $\rho_i$, taken from
the set $\{\rho_0,\rho_1,\dots,\rho_{M-1}\}$ as
$\rho_i=\rho_{i\,{\rm mod}\, M}$. Again, for simplicity, we will
only consider the case of $M=3$ with just two possible
self-transition probabilities, $r_1, \,r_2$, depending only on the
capital being a multiple of three or not:
$r_1\equiv\rho_0,\,r_2\equiv\rho_1=\rho_2$.

As we will show, the significance of this new version is a natural
evolution of Parrondo's games, which can now be rigorously derived
from the Fokker-Planck equation, based on a physical flashing
ratchet model.

The outline of this paper is as follows. In section $2$ we briefly review two
relations concerning Parrondo's games and the Fokker-Planck equation. In both
relations it is straightforward the inclusion of self-probabilities. In Sec.
$3$ we give a mathematical analysis of the new games using discrete-time Markov
chains and derive conditions for the paradox to appear. In Sec. $4$ we
calculate the rates of winning, describe the parameter space and present
numerical simulations which confirm and extend the theoretical analysis.
Finally, in section $5$ we provide a brief discussion of the results.

\section{The Flashing ratchet and the Fokker-Planck equation}
Despite the fact that Parrondo's paradox was inspired by the
flashing ratchet, the relation between both has only been made
quantitative recently, when two different approaches have
established a formal relation between Parrondo's games and the
physical model of the flashing ratchet (Allison \& Abbott 2002 and
Toral \emph{et al.} 2003a). We now very briefly  review both
approaches.

In the scheme proposed by Allison \& Abbott (2002), the starting
point is the following general Fokker--Planck equation (see
Horsthemke and Lefever, 1984), for the probability $P(x,t)$ of a
Brownian particle moving in a time-dependent one-dimensional
potential $V(x,t)$:
\begin{equation} \label{eq21}
D \frac{\partial^2P}{\partial x^2}-P \frac{\partial
\alpha}{\partial x}-\alpha\frac{\partial P}{\partial
x}-\frac{\partial P}{\partial t}= 0,
\end{equation}
where $\alpha$ and $D$ are referred to as the infinitesimal first
and second moments of diffusion, respectively; $D$ has a constant
value -- ``Fick's law constant"-- while $\alpha(x,t)$ is a
function related to the applied potential $V(x,t)$  by the
equation
\begin{equation} \label{eq22}
\alpha(x,t)=-u\frac{\partial}{\partial x}V(x,t),
\end{equation}
where $u$ denotes the mobility of the Brownian particle.

Then (\ref{eq21}) is discretized using a finite difference
approximation obtaining
\begin{equation}
P_{i,j}=a_{-1}^{i,j}\cdot P_{i-1,j-1}+a_{0}^{i,j}\cdot
P_{i,j-1}+a_{+1}^{i,j}\cdot P_{i+1,j-1} \label{eq23}
\end{equation}
where
\begin{equation}
a_{-1}^{i,j}=\frac{\frac{D\tau}{\lambda^2}+\frac{\alpha(i,j)
\tau}{2\lambda}}{\frac{\alpha(i+1,j-1)-\alpha(i-1,j-1)}{2\lambda}\tau+1}
\label{eq24}
\end{equation}
\begin{equation}
a_{0}^{i,j}=\frac{-2\frac{D\tau}{\lambda^2}+1}{\frac{\alpha(i+1,j-1)-\alpha(i-1,j-1)}
{2\lambda}\tau+1} \label{eq25}
\end{equation}
\begin{equation}
a_{+1}^{i,j}=\frac{\frac{D\tau}{\lambda^2}-\frac{\alpha(i,j)
\tau}{2\lambda}}{\frac{\alpha(i+1,j-1)-\alpha(i-1,j-1)}{2\lambda}\tau+1}\,.
\label{eq26}
\end{equation}\\

Here the index $i$ denotes the discretized space $x=i\lambda$,
whereas $j$ denotes the discretized time $t=j\tau$ ; $\lambda$ and
$\tau$ account for the space and time discretization steps
respectively.

This discretized form (\ref{eq23}) of the Fokker--Planck equation
is compared to the master equation for any of the gambling games
used in Parrondo's paradox
\begin{equation}
P_{i,j}=\pi_{i-1}\cdot P_{i-1,j-1}+\rho_i\cdot
P_{i,j-1}+(1-\pi_{i+1}-\rho_{i+1})\cdot P_{i+1,j-1} \label{eq27}
\end{equation}
where $P_{i,j}$ denotes the probability that the player has a
capital $i$ at the $j$ play. In the original Parrondo games the
self-transition probability is zero, so that the term  $\rho_i$ is
set to zero in the following calculations.

Combining  (\ref{eq23}) and  (\ref{eq27}) we get
\begin{equation}
\frac{\pi_{i-1}}{1-\pi_{i+1}}=\frac{a_{-1}^{i,j}}{a_{+1}^{i,j}}=\frac{1+\frac{\lambda}{2D\tau}\alpha(i,j)}{1-\frac{\lambda}{2D\tau}\alpha(i,j)}
\end{equation}
and it follows that the function $\alpha(i,j)\equiv\alpha_i$ is
independent of the time index $j$:
\begin{equation}
\label{alfa}
\alpha_{i}=\frac{2D}{\lambda}\,\frac{\pi_{i-1}-(1-\pi_{i+1})}{\pi_{i-1}+(1-\pi_{i+1})}\,.
\end{equation}
Finally, the discretized values of the potential are obtained
combining (\ref{eq22}) with  (\ref{alfa}),
\begin{equation}
V_i=-\frac{2D}{u}\sum^i_{k=0}\frac{1-\left(\frac{1-\pi_{k+1}}{\pi_{k-1}}\right)}
{1+\left(\frac{1-\pi_{k+1}}{\pi_{k-1}}\right)}.\label{eq28}
\end{equation}
This equation allows one to obtain the discretized version of the
physical potential $V_i$ in terms of the probabilities $\pi_i$
of the games.\\

A second relation between the Fokker--Planck equation and the
master equation has been proposed by Toral \emph{et al.} (2003a).
Unlike the first approach described above now the starting point
is not the Fokker--Planck equation but rather the rewriting of the
master equation (\ref{eq27}) in the form of a continuity equation
for the probability:
\begin{equation}
P_{i,j}-P_{i,j-1}=-\left[J_{i+1,j}-J_{i,j}\right] \label{eq211}
\end{equation}
where the current $J_{i,j}$ is given by:
\begin{equation}
\label{current} J_{i,j}=\frac{1}{2}\left[F_i
P_{i,j}+F_{i-1}P_{i-1,j}\right]-\left[D_iP_{i,j}-D_{i-1}P_{i-1,j}\right]
\end{equation}
and
\begin{equation}
F_i = 2\pi_i+\rho_i-1,\hspace{2.0cm} D_i = \frac{1-\rho_i}{2}\,.
\label{eq213}
\end{equation}

These coefficients can be related with their analogous terms
corresponding to a discretization of the Fokker--Planck equation
for a probability $P(x,t)$
\begin{equation}
\frac{\partial P(x,t)}{\partial t}=-\frac{\partial
J(x,t)}{\partial x} \label{eq214}
\end{equation}
with a current
\begin{equation}
J(x,t)=F(x)P(x,t)-\frac{\partial [D(x)P(x,t)]}{\partial x} \label
{eq215}
\end{equation}
for a general drift $F(x)$ and diffusion $D(x)$.\\

Considering again the case $\rho_i=0$, we have
\begin{equation}
D_i\equiv D=1/2\hspace{2.0truecm} F_i=-1+2\pi_i \label{eq216}
\end{equation}
and the following form for the current:
\begin{equation}\label{eq216a}
J_{i,j}=\pi_{i-1}P_{i-1,j}-(1-\pi_i)P_{i,j}
\end{equation}
which is nothing but the probability flux from state $i-1$ to
state $i$.

The relation between the external potential $V_i$ and the games
probabilities is in this formulation written as:
\begin{equation}
V_i=-\frac{1}{2}\sum_{k=1}^i\ln\left[\frac{1+F_{k-1}}{1-F_k}\right]=
-\frac{1}{2}\sum_{k=1}^i\ln\left[\frac{\pi_{k-1}}{1-\pi_k}\right]
\label{eq217}
\end{equation}
where the value $V_0=0$ has been adopted for convenience. This
equation is the main result concerning the relation between the
games probabilities $\pi_i$ and the discretized version of the
potential $V_i$. As with  (\ref{eq28}), through (\ref{eq217}) we
can obtain the potential that corresponds to a Parrondo game.
Notice that both approaches yield different values for the
potential $V_i$ corresponding to a set of games probabilities
$\{\pi_0,\dots,\pi_{M-1}\}$. For instance, in the case of a fair
game, the potential given by (\ref{eq217}) is a periodic function
$V_{i+M}=V_i$, (Toral \emph{et al.} 2003b). Nevertheless, it can
be shown that both potentials coincide in the limit of an
infinitessimaly small space discretized step $\lambda$.

It is possible to solve the master equation (\ref{eq211}) using a
constant current $J_{i,j}=J$, together with the boundary condition
$P_i^{st}=P_{i+M}^{st}$ in order to obtain the stationary
probability distribution $P_i^{st}$. The result is:
\begin{equation}
P_i^{st}=N {\rm e}^{-2V_i}\left[1- \frac{2 J}{N} \sum_{j=1}^i
\frac{{\rm e}^{2V_j}}{1-F_j}\right] \label{eq218}
\end{equation}
with a current
\begin{equation}
\label{eq219} J=N\frac{{\rm e}^{-2V_M}-1}{2\sum_{j=1}^M \frac{{\rm
e}^{2V_j}}{1-F_j}}
\end{equation}
and $N$ is a normalization constant obtained from
$\sum_{k=0}^{M-1}P_k^{st}=1$.

The inverse problem of obtaining the game probabilities in terms
of the potential can also be done. It requires solving
(\ref{eq217}) with the boundary condition $F_0=F_{M}$. The
solution is given by
\begin{equation}
\label{eq220} F_i=(-1)^i{\rm
e}^{2V_i}\left[\frac{\sum_{j=1}^{M}(-1)^j[{\rm e}^{-2V_j}-{\rm
e}^{-2V_{j-1}}]}{(-1)^M{\rm
e}^{2(V_0-V_M)}-1}+\sum_{j=1}^{i}(-1)^j[{\rm e}^{-2V_j}-{\rm
e}^{-2V_{j-1}}]\right]
\end{equation}
which, via $\pi_i=(1+F_i)/2$, allows one to obtain the
probabilities $\pi_i$ in terms of the potential $V_i$. It is clear
that the additional condition $\pi_i\in[0,1],\,\forall i$ must be
fulfilled by any acceptable solution.

\begin{figure}[!htb]
\centerline{\epsfig{figure=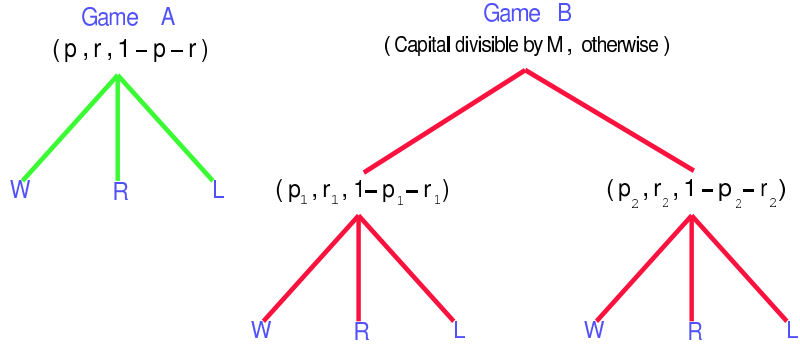,width=10.5cm,height=4.5cm}}
\vspace{0.5truecm} \caption{\label{Fig1} Probability trees of the
new games A and B. Game A is formed by three branches, denoting
the three possibilities of winning (W), remaining in the same
state (R) and losing (L). Note that game B has a capital dependent
rule and therefore is not a martingale.}
\end{figure}

To sum up, we have two approaches, either (\ref{eq28}) or
(\ref{eq217}), that allow one to obtain the potential
corresponding to a set of probabilities $(\pi_0,\dots,\pi_{M-1})$
defining a Parrondo game. In both approaches it is very easy to
introduce self-probabilities $\rho_i\ne 0$. Therefore, we find it
interesting to investigate the effect of these terms in the
Parrondo's paradox. Therefore we introduce a new branch in the
original games (Harmer \& Abbott 2002) that accounts for the
self-transition probability denoted by $r_i$. The new diagrams for
the games A and B are presented in figure \ref{Fig1}. In the next
section we will investigate the effect of this new inclusion upon
the Parrondo effect.

\section{Analysis of the new Parrondo games with self-transitions }
\subsection{Game A}
We start with the new game A, where the probability of winning is
$p$, the probability of remaining with the same capital will be
denoted as $r$, and we lose with probability $q=1-r-p$ .

Following the same reasoning as Harmer \emph{et al.} (2000)  we
will calculate the probability $f_j$ that our capital reaches zero
in a finite number of plays, supposing that initially we have a
given capital of $j$ units. From Markov chain analysis (Karlin
1973) we have that

\begin{itemize}
    \item $f_j=1$ for all $j\ge{0}$, and so the game is either
    fair or losing; or
    \item $f_j<1$ for all $j>0$, in which case the game can be
    winning because there is a certain probability that our capital
    can grow indefinitely.
\end{itemize}

We are looking for the set of numbers $\{f_j\}$ that correspond to
the minimal non-negative solution of the equation

\begin{equation} \label{eq32}
f_j=p\cdot f_{j+1}+r\cdot f_j+q\cdot f_{j-1}\,,\hspace{0.5truecm}
j\geq1
\end{equation}
with the boundary condition
\begin{equation} \label{eq33}
f_{0}=1\,.
\end{equation}
With a subtle rearrangement,  (\ref{eq32}) can be put in the
following form
\begin{equation} \label{eq34}
f_j=\frac{p}{1-r}\cdot f_{j+1}+\frac{q}{1-r}\cdot f_{j-1}\,.
\end{equation}
Whose solution, for the initial condition (\ref{eq33}), is \makebox{$f_j=A
\cdot[(\frac{1-p-r}{p})^j-1] + 1$}, where $A$ is a constant. For
the minimal non-negative solution we obtain
\begin{equation} \label{eq35}
f_j=\min\left[1,\left(\frac{1-p-r}{p}\right)^j\right].
\end{equation}
So we can see that the new game A is a winning game for
\begin{equation}\label{eq35a}
\frac{1-p-r}{p}<1,
\end{equation}
is a losing game  for
\begin{equation}\label{eq35b}
\frac{1-p-r}{p}>1,
\end{equation}
and is a fair game for
\begin{equation}\label{eq35c}
\frac{1-p-r}{p}=1.
\end{equation}
\subsection{Game B}
We now analyze the new game B. Like game A, we have introduced the
probabilities of a self-transition in each state, that is, if the
capital is a multiple of three we have a probability $r_1$ of
remaining in the same state, whereas if the capital is not a
multiple of three then the probability is $r_2$. The rest of the
probabilities will follow the same notation as in the original
game B, so we have the following scheme \\

\begin{equation} \label{eq36}
\left\{
\begin{array}{cc}
           \mathrm{mod(capital,3)}=0 \rightarrow p_1, r_1, q_1 \\
           \\
           \mathrm{mod(capital,3)}\neq 0 \rightarrow p_2, r_2, q_2.
          \end{array}
       \right.
\end{equation}\\

As in the case of game A, we will follow similar reasoning as Harmer
\emph{et al.} (2000) but for game B. Let $g_j$ be the probability
that the capital will reach the zeroth state in a finite number of
plays, supposing an initial capital of $j$ units. Again, from Markov
chain theory we have\\
\begin{itemize}
    \item $g_j=1$ for all $j\ge{0}$, so game B is either
    fair or losing; or
    \item $g_j<1$ for all $j>0$, in which case game B can be
    winning because there is a certain probability for the capital
    to grow indefinitely.
\end{itemize}
For $j\ge{1}$, the following set of recurrence equations must be
solved:
\begin{equation} \label{eq37}
\begin{array}{rcll}
g_{3j  }&=& p_1\cdot g_{3j+1}+r_1\cdot g_{3j}+(1-p_1-r_1)\cdot
g_{3j-1},&\hspace{0.4truecm}j\geq1 \\
g_{3j+1}&=& p_2\cdot g_{3j+2}+r_2\cdot g_{3j+1}+(1-p_2-r_2)
\cdot g_{3j},&\hspace{0.4truecm}j\geq0\\
g_{3j+2}&=& p_2\cdot g_{3j+3}+r_2\cdot g_{3j+2}+(1-p_2-r_2)\cdot
g_{3j+1},&\hspace{0.4truecm}j\geq0\,.
\end{array}
\end{equation}\\
As in game A, we are looking for the set of numbers $\{g_j\}$ that
correspond to the minimal non-negative solution. Eliminating terms
$g_{3j-1}$, $g_{3j+1}$ and $g_{3j+2}$ from (\ref{eq37}) we get
\begin{equation} \label{eq38}
[p_1 p^2_2 + (1-p_1-r_1)(1-p_2-r_2)^2] \cdot g_{3j}=p_1 p^2_2
\cdot g_{3j+3}+(1-p_1-r_1)(1-p_2-r_2)^2 \cdot g_{3j-3}\,.
\end{equation}

Considering the same boundary condition as in game A, $g_0=1$,
the last equation has a general solution of the form $g_{3j}=B
\cdot \left[\left(\frac{(1-p_1-r_1)(1-p_2-r_2)^2}{p_1
p^2_2}\right)^j-1\right]+1$, where $B$ is a constant.  For the
minimal non-negative solution we obtain\\

\begin{equation} \label{eq39}
g_{3j}=\min \left[1,\left(\frac{(1-p_1-r_1)(1-p_2-r_2)^2}{p_1
p^2_2}\right)^j\right].
\end{equation}\\

It can be verified that the same solution (\ref{eq39}) will be
obtained solving (\ref{eq37}) for $g_{3j+1}$ and $g_{3j+2}$,
leading all them to the same condition for the probabilities of
the games.

As with game A, game B will be winning if
\begin{equation}\label{eq39a}
\frac{(1-p_1-r_1)(1-p_2-r_2)^2}{p_1 p^2_2}<1,
\end{equation}
losing if
\begin{equation}\label{eq239b}
\frac{(1-p_1-r_1)(1-p_2-r_2)^2}{p_1 p^2_2}>1,
\end{equation}
and fair if
\begin{equation}\label{eq239c}
\frac{(1-p_1-r_1)(1-p_2-r_2)^2}{p_1 p^2_2}=1.
\end{equation}

\subsection{Game AB}
Now we will turn to  the random alternation of games A and B with
probability $\gamma$. This will be named as game AB. For this game
AB we have the following (primed) probabilities
\begin{itemize}
    \item if the capital is a multiple of three
     \begin{equation} \label{eq310}
        \left\{\begin{array}{cc}
        p'_1=\gamma\cdot p+(1-\gamma)\cdot p_1, \\
        r'_1=\gamma\cdot r+(1-\gamma)\cdot r_1, \\
        \end{array}\right.
     \end{equation}
    \item if the capital is not  multiple of three
      \begin{equation} \label{eq311}
        \left\{\begin{array}{cc}
         p'_2=\gamma\cdot p+(1-\gamma)\cdot p_2, \\
         r'_2=\gamma\cdot r+(1-\gamma)\cdot r_2. \\
        \end{array}\right.
      \end{equation}
\end{itemize}

The same reasoning as with game B can be made but with the new
probabilities $p'_1$, $r'_1$, $p'_2$, $r'_2$ instead of $p_1$,
$r_1$, $p_2$, $r_2$. Eventually we obtain that game AB will be
winning if
\begin{equation} \label{eq312}
\frac{(1-p'_1-r'_1)(1-p'_2-r'_2)^2}{p'_1 p'^2_2}<1,
\end{equation}\\
losing if
\begin{equation} \label{eq313}
\frac{(1-p'_1-r'_1)(1-p'_2-r'_2)^2}{p'_1 p'^2_2}>1,
\end{equation}\\
and fair if
\begin{equation} \label{eq314}
\frac{(1-p'_1-r'_1)(1-p'_2-r'_2)^2}{p'_1 p'^2_2}=1.
\end{equation}\\

The paradox will be present if games A and B are losing, while game AB is
winning. In this framework this means that the conditions (\ref{eq35b}),
(\ref{eq239b}) and (\ref{eq312}) must be satisfied simultaneously. In order to
obtain sets of probabilities fulfilling theses conditions we have first
obtained sets of probabilities yielding {\sl fair} A and B games but such that
AB is a winning game, and then introducing a small biasing parameter $\epsilon$
making game A and game B losing games, but still keeping a winning AB game. As an example, we give some sets of probabilities that fulfill these conditions:
\begin{equation}\begin{array}{rlllllll} \label{probabilitiesa}
(a) & p &= \frac{9}{20}-\epsilon,~  &r=\frac{1}{10},~ &p_1 = \frac{9}{100}-\epsilon,~ &r_1=\frac{1}{10},~
&p_2 = \frac{3}{5}-\epsilon,~ &r_2=\frac{1}{5},\cr
(b) & p&=\frac{9}{20}-\epsilon,~ &r=\frac{1}{10},~
&p_1=\frac{509}{5000}-\epsilon, &r_1=\frac{1}{10},
&p_2=\frac{7}{10}-\epsilon,~ &r_2=\frac{1}{20},\cr
(c)& p&=\frac{9}{20}-\epsilon,~ &r=\frac{1}{10},~
&p_1=\frac{3}{25}-\epsilon,~ &r_1=\frac{2}{5},~
&p_2=\frac{3}{5}-\epsilon,~ &r_2=\frac{1}{10},\cr
(d) & p &= \frac{1}{4}-\epsilon,~ &r=\frac{1}{2},~  &p_1 = \frac{3}{25}-\epsilon,~ &r_1=\frac{2}{5},~ &p_2 = \frac{3}{5}-\epsilon,~ &r_2=\frac{1}{10}.
\end{array}
\end{equation}

\section{Properties of the Games}
\subsection{Rate of winning}
If we consider the capital of a player at play number $n$, $X_n$
modulo $M$, we can perform a Discrete Time Markov Chain (DTMC)
analysis of the games with a state-space $\{0,1,\ldots,M-1\}$
(\emph{c.f.} Harmer \emph{et al.} 2001). For the case of
Parrondo's games we have $M=3$, so the following set of difference
equations for the probability distribution can be obtained (Lee
\emph{et al.} 2002b):
\begin{equation}\label{eq41}
    \begin{array}{cc}
       P_{0,n+1}&= p_2 \cdot P_{2,n}+r_1 \cdot P_{0,n}+q_2 \cdot P_{1,n} \\
       P_{1,n+1}&= p_1 \cdot P_{0,n}+r_2 \cdot P_{1,n}+q_2 \cdot P_{2,n} \\
       P_{2,n+1}&= p_2 \cdot P_{1,n}+r_2 \cdot P_{2,n}+q_1 \cdot P_{0,n} \\
    \end{array}
\end{equation}
which can be put in a matrix form as
$\textbf{P}_{n+1}=\mathbb{T}\cdot\textbf{P}_{n}$, where
\begin{equation}\label{eq42}
\mathbb{T}=\left(\begin{array}{ccc}
  r_1 & q_2 & p_2 \\
  p_1 & r_2 & q_2 \\
  q_1 & p_2 & r_2 \\
\end{array}\right)
\end{equation}
and
\begin{equation}\label{eq43}
\textbf{P}_{n}=\left(
\begin{array}{c}
  P_{0,n} \\
  P_{1,n} \\
  P_{2,n} \\
\end{array}\right).
\end{equation}

In the limiting case where $n\rightarrow\infty$ the system will
tend to a stationary state characterized by
\begin{equation}\label{eq44}
\mathbf{\Pi}=\mathbb{T}\cdot\mathbf{\Pi}
\end{equation}
where $\lim_{n\rightarrow\infty}\mathbf{P}_n=\mathbf{\Pi}$.\\

Solving (\ref{eq44}) is equivalent to solving an eigenvalue problem. As we are
dealing with Markov chains, we know that there will be an eigenvalue
$\lambda=1$ and the rest will be under $1$ (Karlin 1973). For the $\lambda=1$
value we obtain the following eigenvector giving the stationary probability
distribution in terms of the games' probabilities.

\begin{equation}\label{eq45}
\mathbf{\Pi}\equiv\left(
\begin{array}{c}
  \Pi_0 \\
  \Pi_1 \\
  \Pi_2
\end{array}\right)=\frac{1}{D}\left(
\begin{array}{c}
  (1-r_2)^2-p_2 \cdot (1-p_2-r_2) \\
  (1-r_1)(1-r_2)-p_2 \cdot (1-p_1-r_1) \\
  (1-r_1)(1-r_2)-p_1 \cdot (1-p_2-r_2) \\
\end{array}\right)
\end{equation}
where $D$ is a normalization constant given by
\begin{equation}\label{eq45a}
D=(1-r_2)^2+2(1-r_1)(1-r_2)-p_2 (2-p_2-r_2-r_1-p_1)-p_1
(1-p_2-r_2).
\end{equation}

The rate of winning at the $n$--th step, has the general
expression (Harmer \emph{et al.} 2001)
\begin{equation}\label{eq46}
    r(n)\equiv
    E[X_{n+1}]-E[X_n]=\sum^\infty_{i=-\infty}{i\cdot[P_{i,n+1}-P_{i,n}]}\,.
\end{equation}
Using these expressions and by similar techniques to those
employed in Harmer \emph{et al.} (2001) it is possible to obtain
the stationary rate of winning for the new games introduced in the
previous section. The results are, for game A:
\begin{equation}\label{eq49}
    r^{st}_A=2p+r-1
\end{equation}
and for game B
\begin{eqnarray}\label{eq410}
  \nonumber r^{st}_B&=& 2p_2+r_2-1+[q_2-p_2+p_1-q_1]\cdot\Pi_0\\
 & =& \frac{3}{D}\,(p_1 p^2_2-(1-p_1-r_1)(1-p_2-r_2)^2)
\end{eqnarray}
where $D$ is given by (\ref{eq45a}).

It is an easy task to check that when $r_1=r_2=0$ we recover the known
expressions for the original games obtained by Harmer \emph{et al.} (2001). To
obtain the stationary rate for the randomized game AB we just need to replace
in the above expression the probabilities from (\ref{eq310}) and
(\ref{eq311}).

Within this context the paradox appears when $r^{st}_A\le 0$,
$r^{st}_B\le 0$ and $r^{st}_{AB}>0$. If, for example, we use the values from
(\ref{probabilitiesa}d) and a switching probability $\gamma=1/2$, we obtain the
following stationary rates for game A, game B and the random combination AB:

\begin{eqnarray}\label{rateb2}
r^{st}_A & = & -2\epsilon, \nonumber \\
r^{st}_B & = &  \frac{-\epsilon\,(441-120\epsilon+1000\epsilon^2)}{231-40\epsilon+500\epsilon^2},\\
r^{st}_{AB}& =&
\frac{93-9828\epsilon+1920\epsilon^2-32000\epsilon^3}
    {2\,(2499-320\epsilon+8000\epsilon^2)}\,.\nonumber
\end{eqnarray}
which yield the desired paradoxical result for small $\epsilon >0$.

We can also evaluate the stationary rate of winning when both the
probability of winning and the self-transition probability for the
games vary with a parameter $\epsilon$ as $p=p-\frac{\epsilon}{2}$
and $r=r+\epsilon$, so that normalization is preserved. Using the
set of probabilities derived from (\ref{probabilitiesa}d), namely
$p=\frac{1}{4}-\frac{\epsilon}{2}\,,r=\frac{1}{2}+\epsilon\,,p_1=\frac{3}{25}
-\frac{\epsilon}{2}\,,r_1={\frac{2}5}+\epsilon\,,p_2=\frac{3}{5}-\frac{\epsilon}{2}
\,,r_2=\frac{1}{10}+\epsilon$,
the result is:
\begin{eqnarray}\label{rateb1}
r^{st}_A & = & 0,\nonumber\\
r^{st}_B & = &
\frac{-\epsilon\,(21-20\epsilon)}{2\,(77-200\epsilon+125\epsilon^2)},
\\
r^{st}_{AB} & = & \frac{31-164\epsilon+160\epsilon^2}
    {2\,(833-2600\epsilon+2000\epsilon^2)}\,,\nonumber
\end{eqnarray}
again a paradoxical result.

A comparison between the expressions for the rates of winning of
the original Parrondo games (Harmer \emph{et al}. 2001) and the
new games can be done in two ways. The first one consists in
comparing two games with the same probabilities of winning, say
original game A with probabilities $p=\frac{1}{2}$ and
$q=\frac{1}{2}$ and the new game A with probabilities
$p_{\mathrm{new}}=\frac{1}{2}$, $r_{\mathrm{new}}=\frac{1}{4}$ and
$q_{\mathrm{new}}=\frac{1}{4}$. In this case we can think of the
`old' probability of losing $q$ as taking the place of the
\emph{self-transition} probability $r_{\mathrm{new}}$ and the new
probability of losing $q_{\mathrm{new}}$. In this way we obtain a
higher rate of winning in the new game A than in the original game
-- remember that the new game A has an extra term $r$ in the rate
of winning compared to the original rate, and this extra term is
what gives rise to the higher value. The same reasoning applies
for game B, leading to the same conclusion.

The other possibility could be to compare the two games with the
same probability of losing. In this case, we follow the same
reasoning as before, but now we can imagine the `old' probability
of winning as replacing the winning and self-transition
probabilities of the new game. What we now obtain is a lower rate
of winning for the new game compared to the original one. An easy
way of checking this is by rewriting (\ref{eq49}) and
(\ref{eq410}) as
\begin{equation}\label{eq49n}
    r^{st}_A=p-q
\end{equation}
\begin{equation}\label{eq410n}
  \nonumber r^{st}_B= \frac{3}{D}\,(p_1 p^2_2-q_1 q_2^2).
\end{equation}
So for the same value of $q$ but a lower value of $p$ we obtain a
lower value for the rates of game A and B.

\begin{figure}[!htb]
\centerline{\epsfig{figure=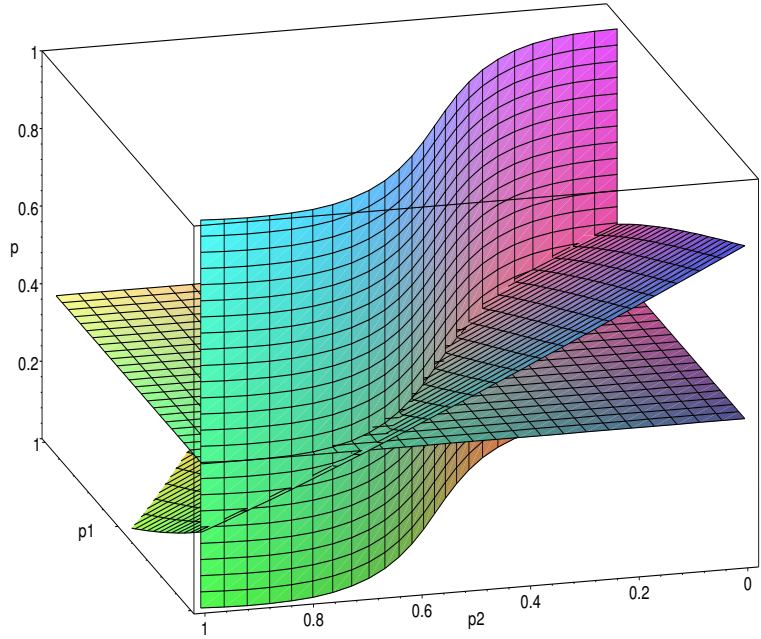,width=10cm,height=8.5cm}}
\vspace{0.5truecm}\caption{\label{Fig2} Parameter space
corresponding to the values $r=\frac{1}{4}$, $r_1=\frac{1}{8}$ and
$r_2=\frac{1}{10}$. The actual region where the paradox exists is
delimited by the plane $p_1=0$ and the triangular region situated
at the frontal face, where all the planes intersect.}
\end{figure}

We now explore the range of probabilities in which the Parrondo
effect takes place. We restrict ourselves to the case $M=3$ and
$\gamma=1/2$ used in the previous formulae.

The fact that we have introduced three new probabilities
complicates the representation of the parameter space as we have
six variables altogether, two variables $\{p,r\}$ from game A and
four variables $\{p_1,r_1,p_2,r_2\}$ coming from game B. In order
to simplify this high number of variables, some probabilities must
be set so that a representation in three dimensions will be
possible. In our case we will fix the variables $\{r,r_1,r_2\}$ so
that the surfaces can be represented in the parameter space
$\{p,p_1,p_2\}$.

In figure \ref{Fig2} we can see the resulting region where the
paradox exists for the variables $r=\frac{1}{4}$,
$r_1=\frac{1}{8}$ and $r_2=\frac{1}{10}$. Some animations have
shown that the volume where the paradox takes place, gradually
shrinks to zero as the variables $r$, $r_1$ and $r_2$ increase
from zero to their maximum value of one.

Another interesting fact that we have encountered, which remains
an open question, is the impossibility of obtaining the equivalent
parameter space to figure \ref{Fig2} with the fixed variables
${p,p_1,p_2}$ and with the parameter space variables ${r,r_1,r_2}$
instead -- it is possible to obtain the planes for games A and B,
but not for the randomized game AB.

\begin{figure}[!htb]
\centerline{\makebox{
\epsfig{figure=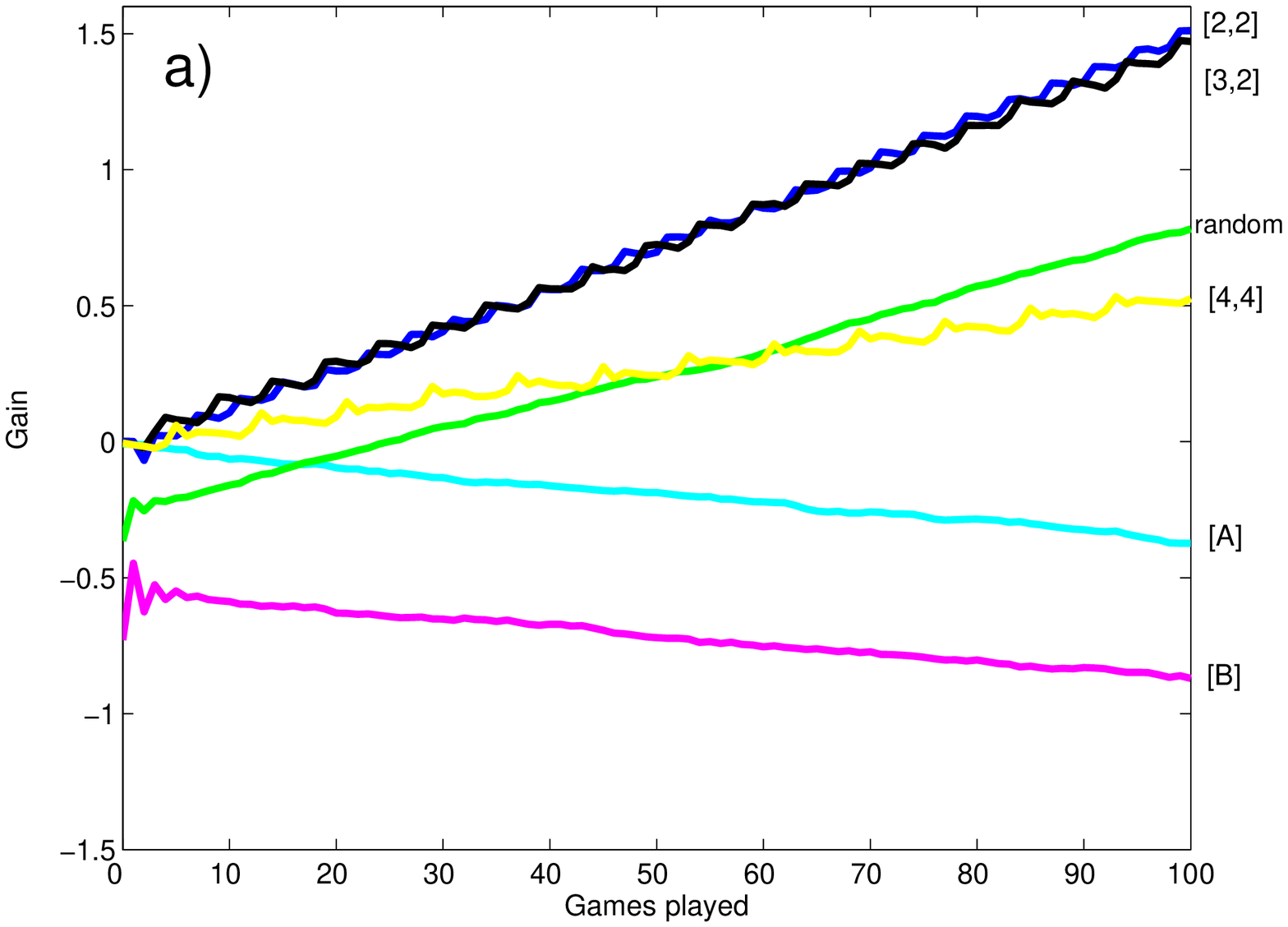,width=6cm,height=5cm}
\epsfig{figure=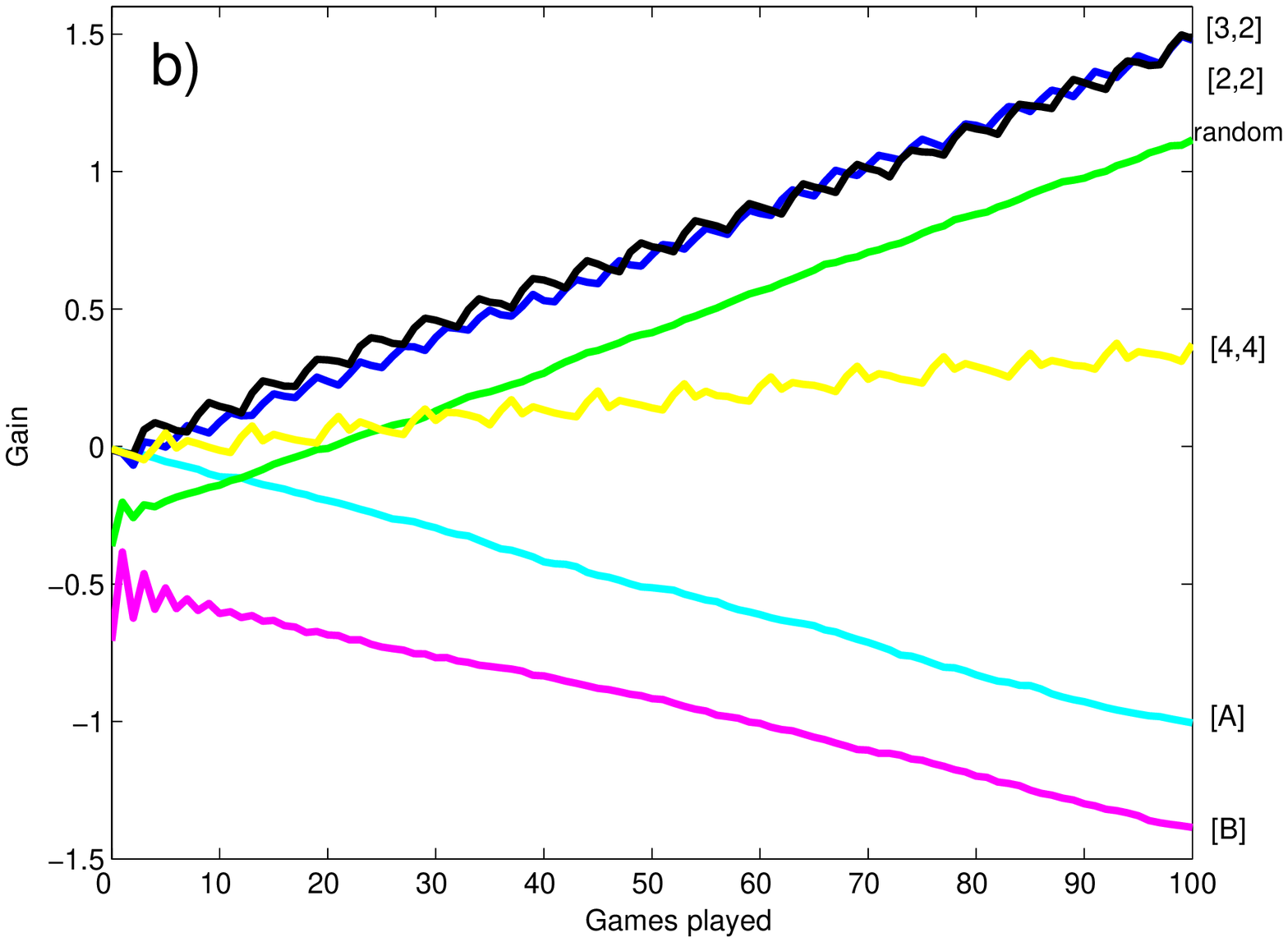,width=6cm,height=5cm}}}
\centerline{\makebox{
\epsfig{figure=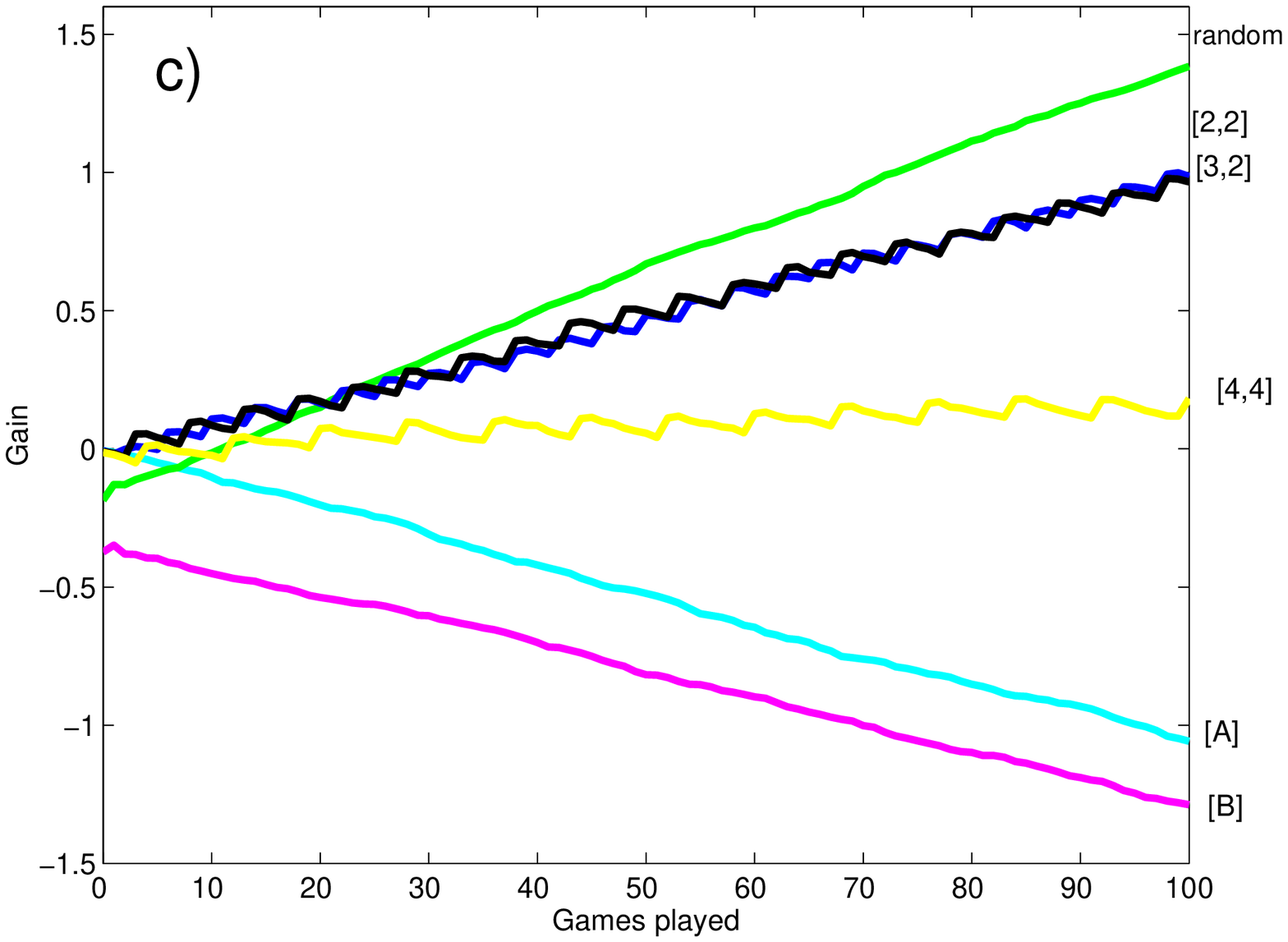,width=6cm,height=5cm}
\epsfig{figure=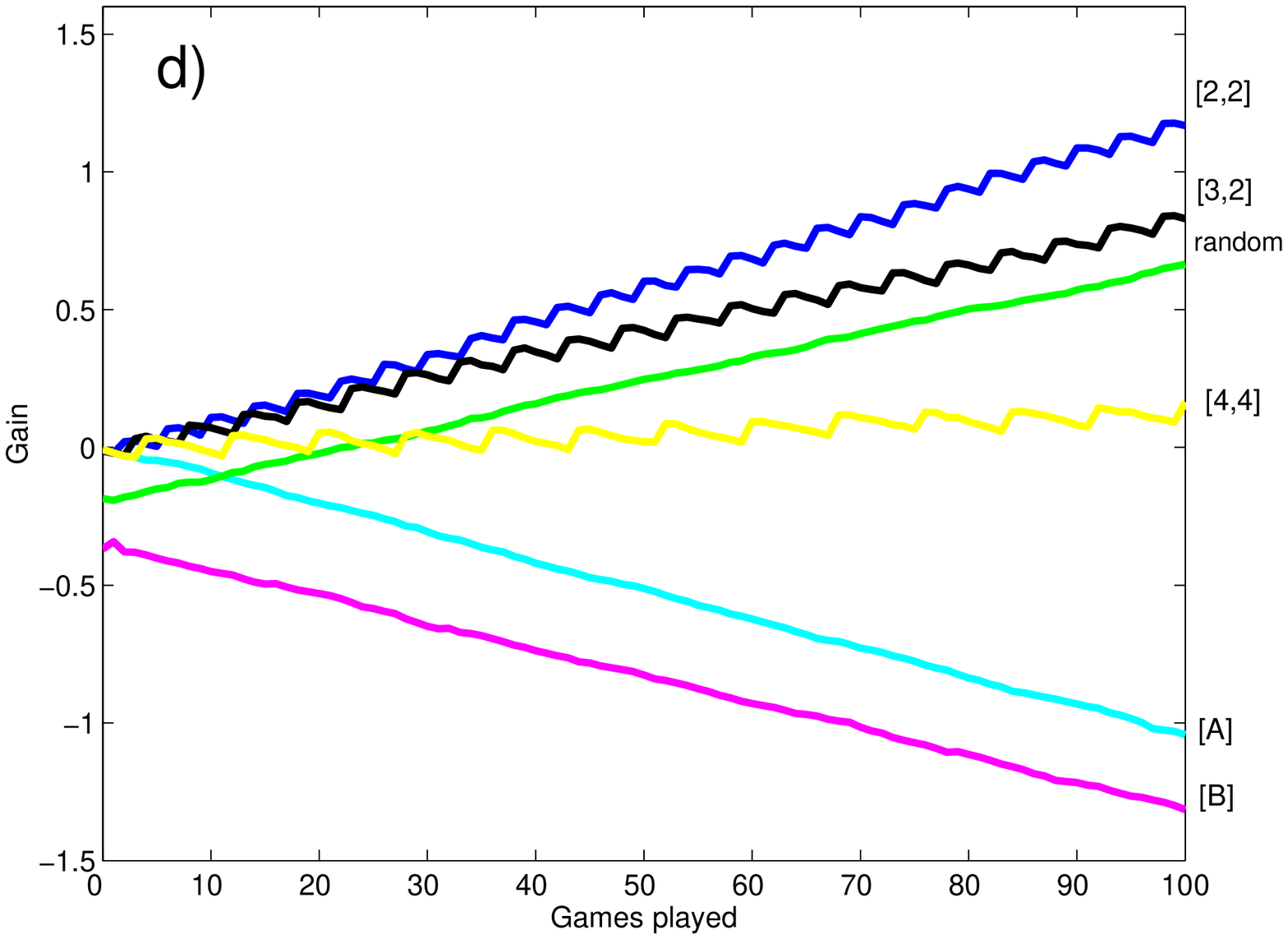,width=6cm,height=5cm}}}
\vspace{0.5truecm} \caption{\label{Fig3} Average gain as a
function of the number of games played coming from numerical
simulation of Parrondo's games with differet sets of
probabilities. The notation $[a,b]$ indicates that game A was
played $a$ times and game B $b$ times. The gains were averaged
over 50\,000 realizations of the games. a) Simulation
corresponding to the probabilities (\ref{probabilitiesa}a) and
$\epsilon=\frac{1}{500}$; b) probabilities (\ref{probabilitiesa}b)
and $\epsilon=\frac{1}{200}$; c) probabilities
(\ref{probabilitiesa}c) and $\epsilon=\frac{1}{200}$; d)
probabilities (\ref{probabilitiesa}d) and $\epsilon=\frac{1}{200}$
. }
\end{figure}

\subsection{Simulations and discussion}
We have analyzed the new games A and B, and obtained the
conditions in order to reproduce the Parrondo effect. We now
present some simulations to verify that the paradox is present for
a different range of probabilities -- see figure \ref{Fig3}. Some
interesting features can be observed from these graphs. First it
can be noticed that the performance of random or deterministic
alternation of the games drastically changes with the parameters.

We use the notation $[a,b]$ to indicate that game A was played $a$
times and game B $b$ times. The performance of the deterministic
alternations $[3,2]$ and $[2,2]$ remain close to one another, as
can be seen in figure \ref{Fig3}. However the alternation $[4,4]$
has a low rate of winning because as we play each game four times,
that causes the dynamics of games A and B to dominate over the
dynamic of alternation, thereby considerably reducing the gain.

\begin{figure}[!thb]
\centerline{\makebox{\epsfig{figure=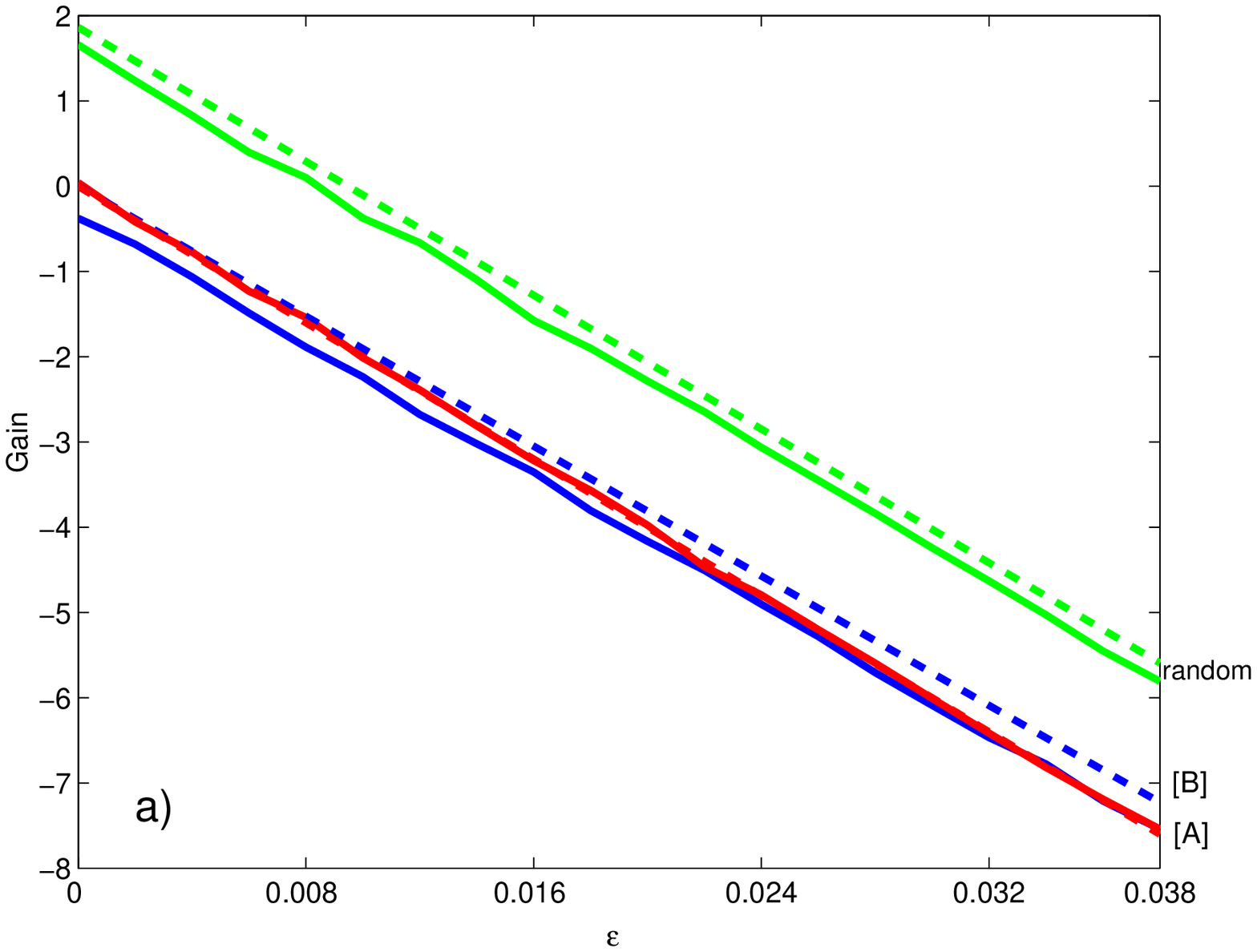,width=6cm,height=5cm}
\epsfig{figure=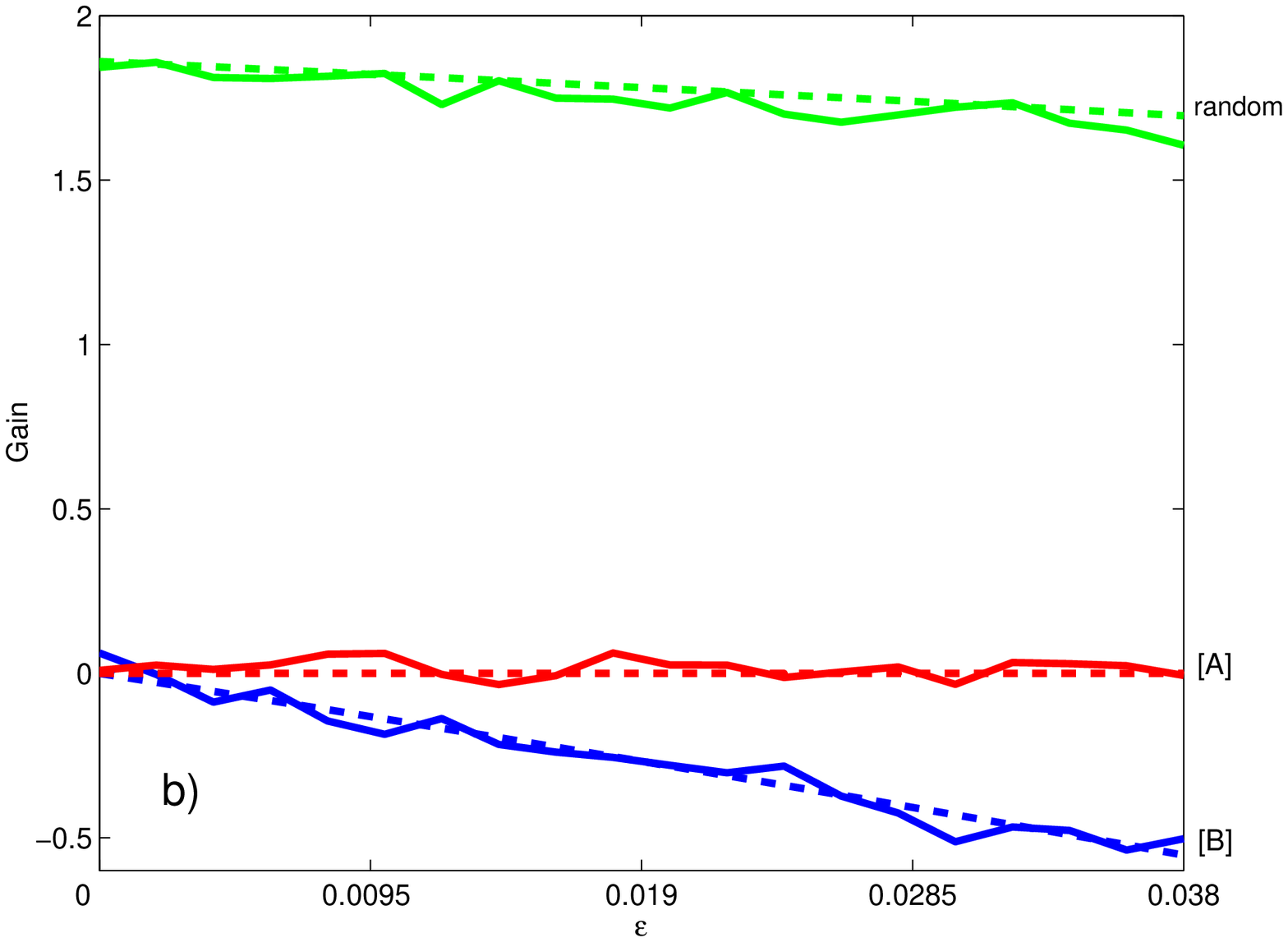,width=6cm,height=5cm}}}
\vspace{0.5truecm}\caption{\label{Fig4} Comparison of the
theoretical rates of winning -- dashed lines -- together with the
rates obtained through simulations -- solid lines. All the
simulations were obtained by averaging over $50\,000$ trials and
over all possible initial conditions. a) The parameters correspond
to the ones used in equations (\ref{rateb2}). b) The parameters
correspond to the ones used in equations (\ref{rateb1}).}
\end{figure}

\begin{figure}[!htb]
\centerline{\epsfig{figure=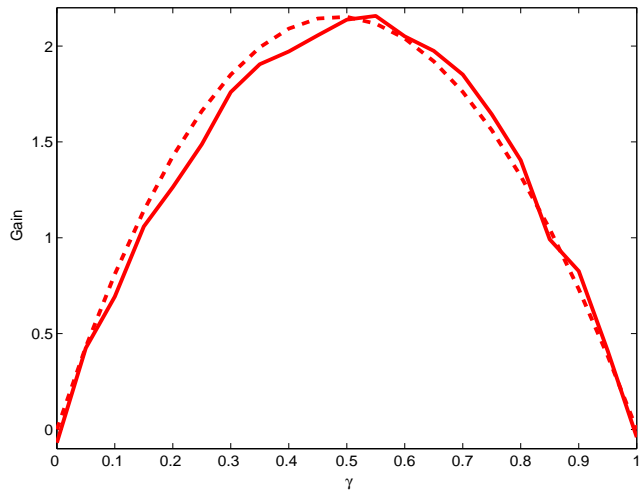,width=8.5cm,height=6.5cm}}
\vspace{0.5truecm} \caption{\label{Fig5} Comparison between the
theoretical and the simulation for the gain vs gamma, for the
following set of probabilities : $p=\frac{1}{3}$, $r=\frac{1}{3}$;
$p_1=\frac{3}{25}$, $r_1=\frac{2}{5}$ and $p_2=\frac{3}{5}$,
$r_2=\frac{1}{10}$. The simulations were carried out by averaging
over $50\,000$ trials and all possible initial conditions.}
\end{figure}

The performance of the random alternation is more variable,
obtaining in some cases a greater gain than in the deterministic
cases -- see  figure \ref{Fig3}c.

In  figures (\ref{Fig4}a) and (\ref{Fig4}b) a comparison between
the theoretical rates of winning for games A, B and AB given by
(\ref{rateb2}) and (\ref{rateb1}) and the rates obtained through
simulations is presented. It is worth noting the good agreement
between both results.

It is also interesting to see how evolves the average gain
obtained from the random alternation of game A and game B when
varying the mixing parameter $\gamma$. In figure \ref{Fig5} we
compare both the experimental and theoretical curves. As in the
original games, the maximum gain is obtained for a value around
$\gamma\sim\frac{1}{2}$ (Lee \emph{et al.} 2002b).

\section{Conclusion}
We have reviewed how the derivation of Parrondo's games from the
flashing Brownian ratchet can be rigorously established via the
Fokker-Planck equation. This procedure reveals new Parrondo games,
of which the original Parrondo games are a special case with
self-transitions set to zero. This confirms Parrondo's original
intuition based on a flashing ratchet is correct with rigour. We
interpreted the self-transitions in terms of particles, in the
flashing ratchet, that remain stationary in a given cycle. We then
presented a new DTMC analysis for the new games showing that
Parrondo's paradox still occurs if the appropriate conditions are
fulfilled. New expressions for the rates of winning have been
obtained, with the result that within certain conditions a higher
rate of winning than in the original games can be obtained. We
have also studied how the parameter space where the paradox exists
changes with the self-transition variables, and conclude that the
parameter space corresponding to the original Parrondo's games is
a limiting case of the maximum volume -- as the self-transition
probabilities increase in value the volume shrinks to zero.
However, it is worth noting that despite the volume
decreases with increasing the self-transition probabilities,
the rates of winning that can be obtained are higher than in the
original Parrondo's games. \\

\footnotesize{This work was supported by GTECH Australasia; the
Ministerio de Ciencia y Tecnolog{\'\i}a (Spain) and FEDER,
projects BFM2001-0341-C02-01 and BFM2000-1108; P.A. acknowledges
support form the Govern Balear, Spain.}

\end{document}